\documentclass[preprint,11pt]{elsarticle}
\usepackage{amsmath}
\usepackage{epsfig}
\usepackage{epstopdf}
\usepackage{subcaption}
\usepackage{graphicx}
\biboptions{numbers,sort&compress}
\usepackage{hyperref}
\hypersetup{
	colorlinks   = true, %Colours links instead of ugly boxes
	urlcolor     = blue, %Colour for external hyperlinks
	linkcolor    = blue, %Colour of internal links
	citecolor   = blue %Colour of citations
}
\usepackage{amssymb}
\begin{document}

\begin{frontmatter}



\title{Nuclear structure investigation of even-even Sn isotopes within the covariant density functional theory}

 \author{Y. EL BASSEM\footnote{younes.elbassem@ced.uca.ma}}%\corref{cor1}}
 %\ead{oulne@uca.ma}
 
 \author{M. OULNE\footnote{oulne@uca.ma}}

 %\cortext[cor1]{Correspondant author}
 \address{High Energy Physics and Astrophysics Laboratory, Department of Physics, \\Faculty of Sciences SEMLALIA, Cadi Ayyad University,  \\P.O.B. 2390, Marrakesh, Morocco.}

\begin{abstract}
	The current investigation aims to study the ground-state properties of one of the most interesting isotopic chains in the periodic table, $^{94-168}$Sn, from the proton drip line to	the neutron drip line by using the covariant density functional theory, which is a modern theoretical tool for the description of nuclear structure phenomena. 
	The physical observables of interest include
	the binding energy, separation energy, two-neutron shell gap, rms-radii for protons and neutrons, pairing energy and quadrupole deformation. 
	The calculations are performed for a wide range of neutron numbers, starting from the proton-rich side up to the neutron-rich one, 
	by using the density-dependent meson-exchange and the density dependent point-coupling effective interactions. The obtained results
	are discussed and compared with available experimental data and with the already existing results of relativistic Mean Field (RMF) model with NL3 functional. The shape phase transition for Sn isotopic chain is also investigated. A reasonable agreement is found between our calculated results and the available experimental data.
\end{abstract}

\begin{keyword}
Ground-state properties, Sn isotopes, covariant density functional theory.

%\PACS code \sep code 21.10.−k \sep 21.10.Dr \sep 21.10.Ft \sep 21.60.−n

\end{keyword}

\end{frontmatter}

% \linenumbers

\section{INTRODUCTION}

In nuclear physics, several approaches have been developed in order to reliably predict the ground state nuclear properties of all nuclei in the periodic table, among them: 
The macroscopic-microscopic models such as Finite Range Droplet Model
(FRDM)~\cite{moller1995}, the macroscopic models such as Bethe-Weizsäcker mass formula~\cite{samanta2002}, and the microscopic models such as the conventional Hartree Fock method~\cite{skyrme1956,decharg1980,bassem2015,bassem2016,bassem2017} with effective density-dependent interactions based on the nonrelativistic kinematics, in which the spin-orbit and the density-dependent interactions are important ingredients, 
and its relativistic analog, the relativistic mean field (RMF) theory~\cite{walecka1974,reinhard1989} based on the relativistic kinematics, in which the nucleons, the mesons and their interactions are the ingredients. 
The RMF theory has been the subject of considerable attention because of its successful description of many properties of nuclei either in the valley of stability or far away from it. It provides an accurate description of ground-state properties and collective excitation of atomic nuclei~\cite{meng1998,zhou2003,meng1999,geng2004,ring1996,meng1996,ginocchio1997}.

Density functional theories (DFT’s) are extremely useful in order to understand the nuclear many-body dynamics. This is done in terms of the energy density functionals (EDFs), which are approximated by the self-consistent mean-field models. They have been applied with great success in Coulombic systems~\cite{kohn1965,kohn1965(2)}, where the functional can be derived directly from the Coulomb interaction without any phenomenological adjustments. In nuclear physics, the situation becomes much more complicated since the spin and isospin degrees of freedom play an important role and cannot be neglected.

One of the most attractive nuclear DFTs is the covariant density functional theory (CDFT)~\cite{niki2008,niki2014,lalazissis2005,rocamaza2011,niki2002} based on the energy density functionals (EDFs), and is very successful in the nuclear structure analysis ~\cite{meng2015,matev2007,afanasjev2008} as well as in describing very well the ground and excited states throughout the nucleic chart~\cite{abusara2012,agbemava2015,agbemava2017}. An example of application of the CDFT on Cd chain is given in Refs.~\cite{Zhao2014,Zhang2014}.

Tin (Sn, Z=50) isotopes have received huge attention not only because they have a magic proton number (Z=50), but also due to other reasons, including the special interest of Sn isotopes for nuclear astrophysics, and its extremely long isotopic chain. In addition, the two magic neutron numbers, N=50 and N=82, make this isotopic chain especially important to test the deformations and the pairing correlations in a microscopic model~\cite{geng2004}.

In this study, we present a detailed investigation of even-even Sn isotopes for a wide range of neutron numbers, N=44-118, within the CDFT framework by using two of the most recent functionals: The density-dependent meson-exchange DD-ME2~\cite{niki2008,niki2014} and the density-dependent point-coupling DD-PC1~\cite{niki2014,lalazissis2005}, which provide a complete and precise description of different
ground states and excited states over the entire nucleic chart~\cite{karim2015,afanasjev2010}.

This paper is organized as follows: In Section \ref{Theoretical Framework}, the theoretical framework and details of the numerical calculations are presented. The results of the investigations of the ground-state properties such as binding energy,  separation energy,  two-neutron shell gap,  rms-radii for protons and neutrons,  pairing energy and  quadrupole deformation, are presented and discussed in Section \ref{Results and Discussion}.
Finally, a summary of the present study is given in Section ~\ref{Conclusion}.

\section{Theoretical Framework}
\label{Theoretical Framework}
Throughout this paper, two classes of covariant density functional models are used: The density-dependent meson-exchange (DD-ME) model, and the density-dependent point-coupling
(DD-PC) model. 
The main differences between these two models reside in the treatment of the interaction range.
DD-ME has a finite interaction range, while DD-PC uses a zero-range interaction with one additional gradient term in the scalar-isoscalar channel. In this investigation, we use the very successful density-dependent meson-exchange DD-ME2~\cite{lalazissis2005} and density-dependent point-coupling DD-PC1~\cite{niki2008} parameter sets.

\subsection{The meson-exchange model}
In the framework of the meson-exchange model, the nucleus is described as a system of Dirac nucleons interacting via the exchange of mesons with finite masses leading to finite-range interactions~\cite{typel1999,lalazissis2009}.
The isoscalar-scalar $\sigma$ meson, the isoscalar-vector $\omega$ meson, and the isovector-vector $\rho$ meson build the minimal set of meson fields for a quantitative description of nuclei.
The meson-exchange model is defined by the standard Lagrangian density with medium dependence vertices~\cite{gambhir1990}:

\begin{eqnarray}
	\mathcal{L}  &  =\bar{\psi}\left[
	\gamma(i\partial-g_{\omega}\omega-g_{\rho
	}\vec{\rho}\,\vec{\tau}-eA)-m-g_{\sigma}\sigma\right]  \psi\nonumber\\
	&  +\frac{1}{2}(\partial\sigma)^{2}-\frac{1}{2}m_{\sigma}^{2}\sigma^{2}%
	-\frac{1}{4}\Omega_{\mu\nu}\Omega^{\mu\nu}+\frac{1}{2}m_{\omega}^{2}\omega
	^{2}\label{lagrangian}\\
	&  -\frac{1}{4}{\vec{R}}_{\mu\nu}{\vec{R}}^{\mu\nu}+\frac{1}{2}m_{\rho}%
	^{2}\vec{\rho}^{\,2}-\frac{1}{4}F_{\mu\nu}F^{\mu\nu}\nonumber
\end{eqnarray}
where $\psi$ denotes the Dirac spinors and $m$ is the bare nucleon mass. $e$ is the charge of the proton and it vanishes for neutrons. $m_\sigma$ , $m_\omega$, and $m_\rho$ are the masses of $\sigma$ meson, $\omega$ meson, and $\rho$ meson, respectively, 
with the corresponding coupling constants for the mesons to the nucleons as $g_\sigma$ , $g_\omega$, and $g_\rho$, respectively. These coupling constants and unknown meson masses are the Lagrangian equation (\ref{lagrangian}) parameters. $\Omega^{\mu\nu}$, ${\vec{R}}^{\mu\nu}$ and $F^{\mu\nu}$ are the field tensors of the vector fields $\omega$, $\rho$ and the proton:
\begin{equation}
	\Omega^{\mu\nu}=\partial^{\mu}\omega^{\nu}-\partial^{\nu}\omega^{\mu},
\end{equation}
\begin{equation}
	{\vec{R}}^{\mu\nu}=\partial^{\mu}{\vec{\rho}}^{\nu}-\partial^{\nu}{\vec{\rho}}^{\mu},
\end{equation}
\begin{equation}
	F^{\mu\nu}=\partial^{\mu}A^{\nu}-\partial^{\nu}A^{\mu}.
\end{equation}

This linear model has first been introduced by Walecka~\cite{walecka1974}. However,
this simple model does not provide a quantitative description of nuclear system ~\cite{boguta1977,Pannert1987} with interaction terms that are only linear in the meson fields.
In particular, the resulting incompressibility of infinite nuclear matter is much too large~\cite{boguta1977} and nuclear deformations are too small~\cite{gambhir1990}. Therefore,
For a realistic description of complex nuclear system properties
one can either introduce a nonlinear self-coupling or a density
dependence in the coupling constants.

For the nonlinear self-coupling, one has to add the following
term to the Lagrangian:

\begin{equation}
	U(\sigma)~=~\frac{1}{2}m_{\sigma}^{2}\sigma^{2}+\frac{1}{3}g_{2}\sigma
	^{3}+\frac{1}{4}g_{3}\sigma^{4}.
\end{equation}
This model has been successfully used in a number of studies~\cite{reinhard1986,lalazissis1997,toddrutel2005}.

In the case of the density dependent coupling constants one defines the dependence as
\begin{equation}
	g_{i}(\rho) = g_i(\rho_{\rm sat})f_i(x) \quad {\rm for} \quad i=\sigma, \omega, \rho
\end{equation}
$i$ can be any of the three mesons $\sigma$, $\omega$ and $\rho$.
There are no nonlinear terms in the $\sigma$ meson, i.e. $g_2 = g_3 =0$.
The density dependence is given by
\begin{equation}\label{fx}
	f_i(x)=a_i\frac{1+b_i(x+d_i)^2}{1+c_i(x+d_i)^2}.
\end{equation}
for $\sigma$ and $\omega$, and for the $\rho$ meson by
\begin{equation}
	f_\rho(x)=\exp(-a_\rho(x-1)).
\end{equation}
\textit{x} is defined as the ratio between the baryonic
density $\rho$ at a specific location and the baryonic density at saturation
$\rho_{\rm sat}$ in symmetric nuclear matter. The parameters in Eq.\ (\ref{fx})
are constrained as follows:
\begin{equation}
	f_i(1)=1,\,\,\,\, f_{\sigma}^{''}(1)=f_{\omega}^{''}(1),\,\, and \,\, f_{i}^{''}(0)=0.
\end{equation}
These constraints reduce the number of independent parameters for density dependence to
three. 

The present study uses the very successful density-dependent meson-exchange relativistic energy functional DD-ME2~\cite{DD-ME2}  with the parameter
set given in Table \ref{tab1}.

%%%%%%%%%%%%%%%%%%%%%%%%%%%%%%%%%%%%%%%%%%%%%%%%%
\begin{table}[ht]
	\caption{ The parameters of the  DD-ME2 parameterization
		of the Lagrangian. 
		\label{tab1}}
	\begin{center}%
		\begin{tabular}
			[c]		{@{\hspace{0pt}}c@{\hspace{24pt}}@{\hspace{24pt}}c@{\hspace{0pt}}}
			\hline Parameter &  DD-ME2 \\\hline
			$m$   & 939  \\
			$m_{\sigma}$  & 550.1238 \\
			$m_{\omega}$  & 783.000  \\
			$m_{\rho}$    & 763.000 \\
			$g_{\sigma}$  &  10.5396\\
			$g_{\omega}$  & 13.0189\\
			$g_{\rho}$    & 3.6836  \\
			$a_{\sigma}$  & 1.3881 \\
			$b_{\sigma}$  & 1.0943 \\
			$c_{\sigma}$  & 1.7057 \\
			$d_{\sigma}$  & 0.4421 \\
			$a_{\omega}$  & 1.3892 \\
			$b_{\omega}$  & 0.9240 \\
			$c_{\omega}$  & 1.4620 \\
			$d_{\omega}$  & 0.4775 \\
			$a_{\rho}$    & 0.5647 \\
			\hline
		\end{tabular}
	\end{center}
\end{table}
%%%%%%%%%%%%%%%%%%%%%%%%%%%%%%%%%%%%%%%%%%%%%%%%%

\subsection{Point-coupling model}
The Lagrangian for the density-dependent point coupling model~\cite{niki2008,nikolaus1992} is given by:\\

\begin{eqnarray}
	\mathcal{L}  &  =\bar{\psi}\left(i\gamma \cdot \partial-m\right)  \psi\nonumber
	-\frac{1}{2}\alpha_S(\hat\rho)\left(\bar{\psi}\psi\right)\left(\bar{\psi}\psi\right)\\
	&-\frac{1}{2}\alpha_V(\hat\rho)\left(\bar{\psi}\gamma^{\mu}\psi\right)\left(\bar{\psi}\gamma_{\mu}\psi\right)
	-\frac{1}{2}\alpha_{TV}(\hat\rho)\left(\bar{\psi}\vec\tau\gamma^{\mu}\psi\right)\left(\bar{\psi}\vec\tau\gamma_{\mu}\psi\right)\label{Lag-pc}\\
	&-\frac{1}{2}\delta_S\left(\partial_v\bar{\psi}\psi\right)\left(\partial^v\bar{\psi}\psi\right) - e\bar\psi\gamma \cdot A
	\frac{(1 - \tau_3)}{2}\psi \nonumber.
\end{eqnarray}

It contains the free-nucleon Lagrangian, the point coupling interaction terms,
and the coupling of the proton to the electromagnetic field. The derivative terms
in Eq.\ (\ref{Lag-pc}) account for the leading effects of finite-range interaction
which are important in nuclei. In analogy with DD-ME models, this model contains isoscalar-scalar, isoscalar-vector and isovector-vector interactions. 
In this work, we have used the recently developed density-dependent point-coupling interaction DD-PC1~\cite{niki2008} given in Table \ref{tab2}. 

%%%%%%%%%%%%%%%%%%%%%%%%%%%%%%%%%%%%%%%%%%%%%%%%%%
\begin{table}[ht]
	\caption{The parameters of the DD-PC1 parameterization in the Lagrangian}%
	\label{tab2}
	\begin{center}%
		\begin{tabular}
			[c]		{@{\hspace{0pt}}c@{\hspace{24pt}}@{\hspace{24pt}}c@{\hspace{0pt}}}
			\hline Parameter &  DD-PC1 \\\hline
			$m$ &   939\\
			$a_{\sigma}$ & -10.04616\\
			$b_{\sigma}$ & -9.15042\\
			$c_{\sigma}$ & -6.42729\\
			$d_{\sigma}$ &  1.37235\\
			$a_{\omega}$ &  5.91946\\
			$b_{\omega}$ &  8.86370\\
			$d_{\omega}$ &  0.65835\\
			$b_{\rho}$   &  1.83595\\
			$d_{\rho}$   &  0.64025\\
			\hline
		\end{tabular}
	\end{center}
\end{table}
%%%%%%%%%%%%%%%%%%%%%%%%%%%%%%%%%%%%%%%%%%%%%%%%%%%%

\section{Results and Discussion}
\label{Results and Discussion}

The numerical results of the ground-state properties of Sn isotopes, such as the binding energy,  separation energy,  two-neutron shell gap,  rms-radii for protons and neutrons,  pairing energy and  quadrupole deformation are presented in this section. In all our calculations, we have used 
two of the state-of-the-art density-dependent effective interactions, namely: the density-dependent meson-exchange DD-ME2~\cite{niki2008} and the density-dependent point-coupling DD-PC1~\cite{lalazissis2005}. Our results are compared with the predictions of the Relativistic Mean Field (RMF) model with NL3~\cite{lalazissis1999} functional and with the available experimental data. We note that the existing experimental data does not cover the full range of our study. Therefore, our results for Sn isotopes, that do not have experimental data, are mainly predictions of future experimental data.\\

\subsection{Binding energy}

Binding energy (BE) is one of the key observables for understanding the structure of a nucleus, and it is directly related to the stability of nuclei. 

\begin{figure}[ht]
	\centering \includegraphics[scale=0.6]{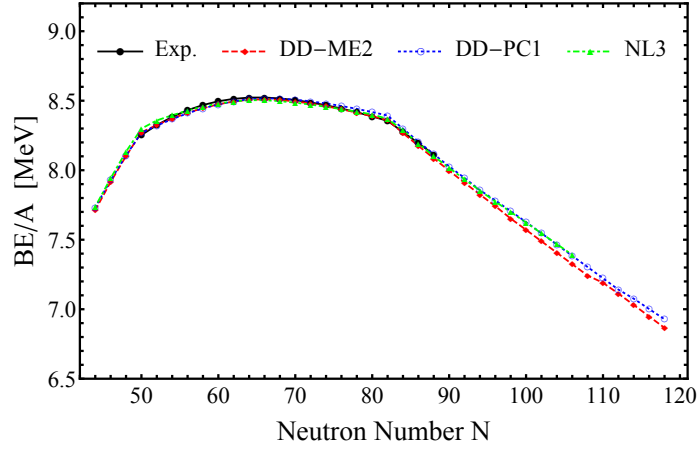}
	\caption{(Color online) The binding energies per nucleon for even-even $^{94-168}$Sn isotopes.}
	\label{BEexp}
\end{figure}

Fig.~\ref{BEexp} shows the binding energies per nucleon (BE/A) for tin isotopes, $^{94-168}$Sn as a function of the neutron number N. The predictions of the RMF model with NL3~\cite{lalazissis1999} functional 
as well as the available experimental data\cite{Audi2003,wang2012} are also shown for comparison.

%%%%%%%%%%%%%%%%%%%%%%%%%%%%%%%%%%%%%%%%%%%%%%%%%%%%%
%%%%%%%%%%%%%%%%%%%%%%%%%%%%%%%%%%%%%%%%%%%%%%%%%%%%%

From Fig. \ref{BEexp}, it is seen that the experimental data are accurately reproduced by the theoretical predictions. 

\begin{figure}[!htb]
	\centering \includegraphics[scale=0.6]{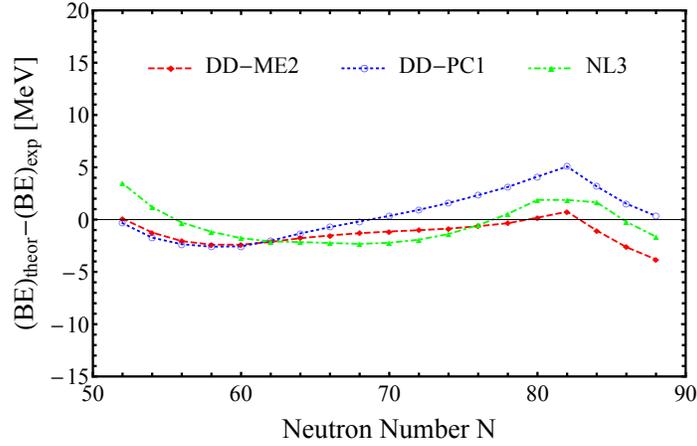}
	\caption{(Color online) Differences between theoretical and experimental total binding energies for even-even Sn isotopes.}
	\label{Delta_BE}
\end{figure}

The differences between the experimental total BE for even-even tin isotopes and the calculated results obtained in this work are shown  in Fig.~\ref{Delta_BE}, as a function of the neutron number N, in order to show to what extent our results are accurate. We emphasize that this comparison is made only for isotopes that have available experimental data.

%%%%%%%%%%%%%%%%%%%%%%%%%%%%%%%%%%%%%%%%%%%%%%%%%%%%%
%%%%%%%%%%%%%%%%%%%%%%%%%%%%%%%%%%%%%%%%%%%%%%%%%%%%%

The comparison between the calculated results and the available experimental total binding energies of tin isotopes is also done by the root mean square (rms) deviation tabulated in Table~\ref{rms}.

%%%%%%%%%%%%%%%%%%%%%%%%%%%%%%%%%%%%%%%%%%%%%%%%%%%%%
\begin{table}[!htb]
	\caption{The rms deviations of the total binding energies of tin isotopes.\label{rms}}
	\centering
	\begin{tabular}
		{@{\hspace{0pt}}c@{\hspace{24pt}}c@{\hspace{24pt}}c@{\hspace{0pt}}}
		\hline\noalign{\smallskip}
		DD-ME2 	&   DD-PC1	&  NL3		 \\ 
		\noalign{\smallskip}\hline\noalign{\smallskip}
		1.7136 	&	2.3180	&	1.7937	 \\  
		\noalign{\smallskip}\hline
	\end{tabular}
\end{table}
%%%%%%%%%%%%%%%%%%%%%%%%%%%%%%%%%%%%%%%%%%%%%%%%%%%%%

From Table~\ref{rms}, one can see that DD-ME2 exceeds in accuracy the other nuclear models DD-PC1 and NL3.

\subsection{Two neutron separation energy ($S_{2n}$) and shell gap ($\delta_{2n}$)}
The neutron  separation energy is an important quantity in testing the validity of a model and in investigating the nuclear shell structure. In this work, we have calculated the two-neutron separation energies, 
$S_{2n}(N,Z)$ = BE$(N,Z)$ - BE$(N - 2,Z)$, for Sn isotopes by using the density-dependent effective interactions DD-ME2 and DD-PC1. 

\begin{figure}[ht]
	\centering \includegraphics[scale=0.6]{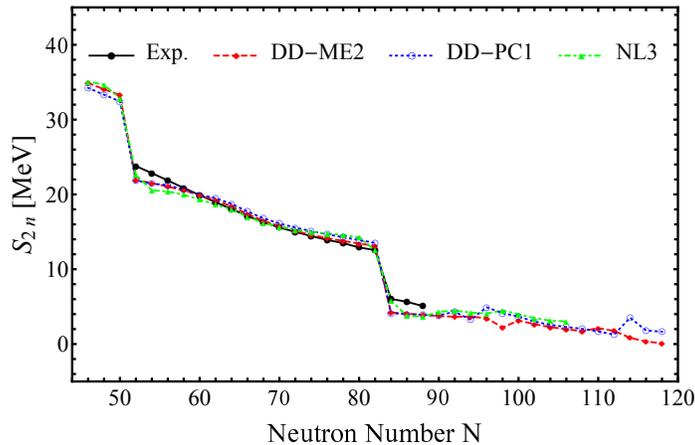}
	\caption{(Color online) The two-neutron separation energies S$_{2n}$, for $^{94-168}$Sn
		isotopes, obtained with  DD-ME2 and DD-PC1, and compared with  NL3
		force~\cite{lalazissis1999} and with the available experimental data~\cite{Audi2003,wang2012}.}
	\label{S2n}
\end{figure}

The calculated two-neutron separation energies S$_{2n}$ of  even-even tin isotopes, as a function of the neutron number N from the proton drip line to the neutron one, are shown in Fig.~\ref{S2n} in comparison with the available experimental data~\cite{Audi2003,wang2012} and with the predictions of RMF(NL3)~\cite{lalazissis1999}.

%%%%%%%%%%%%%%%%%%%%%%%%%%%%%%%%%%%%%%%%%%%%%%%%%%%%%
%%%%%%%%%%%%%%%%%%%%%%%%%%%%%%%%%%%%%%%%%%%%%%%%%%%%%

From Fig.~\ref{S2n}, it is clearly seen that the results of the two density-dependent models as well as those of NL3 reproduce the experimental data quite well. S$_{2n}$ gradually decrease with N, and a strong drop is clearly seen at N = 50 and N = 82 in both experimental and theoretical curves, which indicates the closed shell at these magic neutron numbers. Moreover, as we can see from Table~\ref{rms_S2n}, S$_{2n}$ values obtained by using the density-dependent meson-exchange DD-ME2 are in good agreement with the experimental data and have the lowest rms deviation in comparison with the results obtained within DD-PC1 and NL3 functionals. 

%%%%%%%%%%%%%%%%%%%%%%%%%%%%%%%%%%%%%%%%%%%%%%%%%%%%%
\begin{table}[!htb]
	\caption{The rms deviations of the two-neutron separation energies of tin isotopes.\label{rms_S2n}}
	\centering
	\begin{tabular}
		{@{\hspace{0pt}}c@{\hspace{24pt}}c@{\hspace{24pt}}c@{\hspace{0pt}}}
		\hline\noalign{\smallskip}
		DD-ME2 	&   DD-PC1	&  NL3		 \\ 
		\noalign{\smallskip}\hline\noalign{\smallskip}
		0.8667 	&	1.0121	&	1.0166	 \\  
		\noalign{\smallskip}\hline
	\end{tabular}
\end{table}
%%%%%%%%%%%%%%%%%%%%%%%%%%%%%%%%%%%%%%%%%%%%%%%%%%%%%

\begin{figure}[ht]
	\centering \includegraphics[scale=0.6]{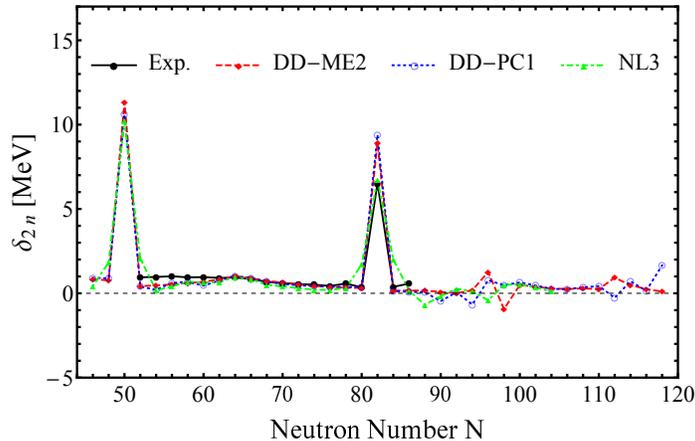}
	\caption{(Color online) The two-neutron shell gap $\delta_{2n}$, for even-even $^{94-168}$Sn isotopes, obtained with  DD-ME2 and DD-PC1, and compared with  NL3
		force~\cite{lalazissis1999} and with the available experimental data~\cite{Audi2003,wang2012}.}
	\label{ShellGap}
\end{figure}

The two-neutron shell gap $\delta_{2n}=S_{2n}(N,Z)-S_{2n}(N+2,Z)$, also known as the S$_{2n}$ differential, is a more sensitive observable for locating the shell closure. The behavior of the $S_{2n}$ differential as a function of the neutron number N is shown in Fig.~\ref{ShellGap}. The sharp peaking in ($\delta_{2n}$) clearly seen at N=50 and N=82 also supports the shell closure at these two neutron magic numbers.

%%%%%%%%%%%%%%%%%%%%%%%%%%%%%%%%%%%%%%%%%%%%%%%%%%%%%
%%%%%%%%%%%%%%%%%%%%%%%%%%%%%%%%%%%%%%%%%%%%%%%%%%%%%

\subsection{Pairing energy ($E_{pair}$)}

\begin{figure}[ht]
	\centering \includegraphics[scale=0.6]{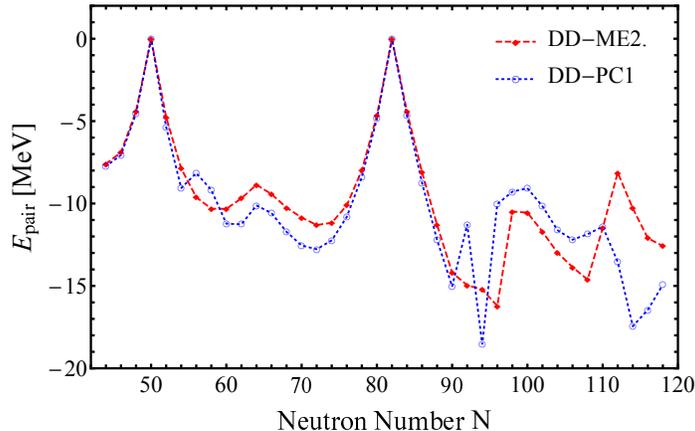}
	\caption{(Color online) The pairing energy for even-even $^{94-168}$Sn isotopes obtained in the CDFT model with DD-ME2 and DD-PC1 interactions.}
	\label{PairingEnergy}
\end{figure}

The pairing energy, defined as the sum of the neutrons and protons contributions, is an important factor that determines the binding energies, masses of nuclei, etc. 
Fig.~\ref{PairingEnergy} shows the pairing energy $E_{pair}$ for $^{94-168}$Sn isotopes calculated by CDFT with the effective interactions DD-ME2 and DD-PC1. 

%%%%%%%%%%%%%%%%%%%%%%%%%%%%%%%%%%%%%%%%%%%%%%%%%%%%%
%%%%%%%%%%%%%%%%%%%%%%%%%%%%%%%%%%%%%%%%%%%%%%%%%%%%%

As one can see from Fig.~\ref{PairingEnergy}, there is an agreement between the results of the two different density-dependent
effective interactions DD-ME2 and DD-PC1. In both cases, $E_{pair}$ vanishes at N = 50 and at N = 82, which confirms the shell closure at these two neutron magic numbers.

\subsection{Quadrupole deformation} 

\begin{figure}[ht]
	\centering \includegraphics[scale=0.6]{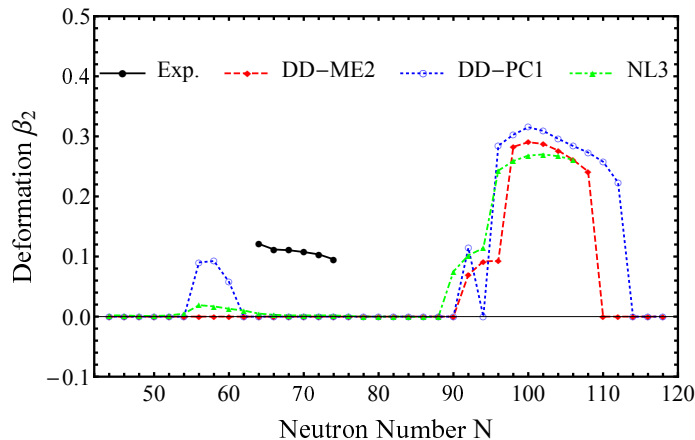}
	\caption{(Color online) The quadrupole deformation parameters, $\beta_2$, for Sn isotopes.}
	\label{beta}
\end{figure}

The quadrupole deformation plays a crucial role in determining some properties of the nuclei such as quadrupole moment, nuclear sizes and isotope shifts.
The quadrupole deformation parameter, $\beta_2$, for Sn isotopes are plotted in Fig.~\ref{beta}. Our results obtained with DD-ME2 and DD-PC1 parameter
sets are compared with those of RMF(NL3)~\cite{lalazissis1999} and with the available experimental data~\cite{beta_experimenta_data}.

%%%%%%%%%%%%%%%%%%%%%%%%%%%%%%%%%%%%%%%%%%%%%%%%%%%%%
%%%%%%%%%%%%%%%%%%%%%%%%%%%%%%%%%%%%%%%%%%%%%%%%%%%%%

From Fig.~\ref{beta}, one can see that the agreement between different theoretical calculations, DD-ME2, DD-PC1 and NL3, is pretty good in general.
Sn isotopes remain robustly spherical till N reaches 90 except for some nuclei where the deformation is quite small.
Nuclei above this neutron number show an interesting change of shape.
It is remarkable that, despite the semi-magic character
of Sn isotopes (Z=50), all the theoretical models predict a significant prolate deformation in the neutron rich region $90 < N < 110$.
The prolate deformation increases and then saturates at a value close to $\beta_2 \approx 0.3$. For nuclei above N=110, there is a transition from deformed to spherical shape.

We would like to make a comment about the  discrepancy seen in Fig.~\ref{beta} between theory and experiment. In fact, the weak deformation exhibited in the available experimental data for even-even Sn isotopes from $N =64$ up to $N =74$ 
is due to the fact that the experimental $\beta_2$ values are usually extracted from experimental $B(E2)$$\uparrow$ values (the reduced probability of the transition from the ground state of the nucleus to the first excited $2^+$ state) by using the Bohr model, which is valid only for well-deformed nuclei \cite{Zongqiang2012}.

\begin{figure}[!ht]
	\centering \includegraphics[width=12cm,height=14cm]{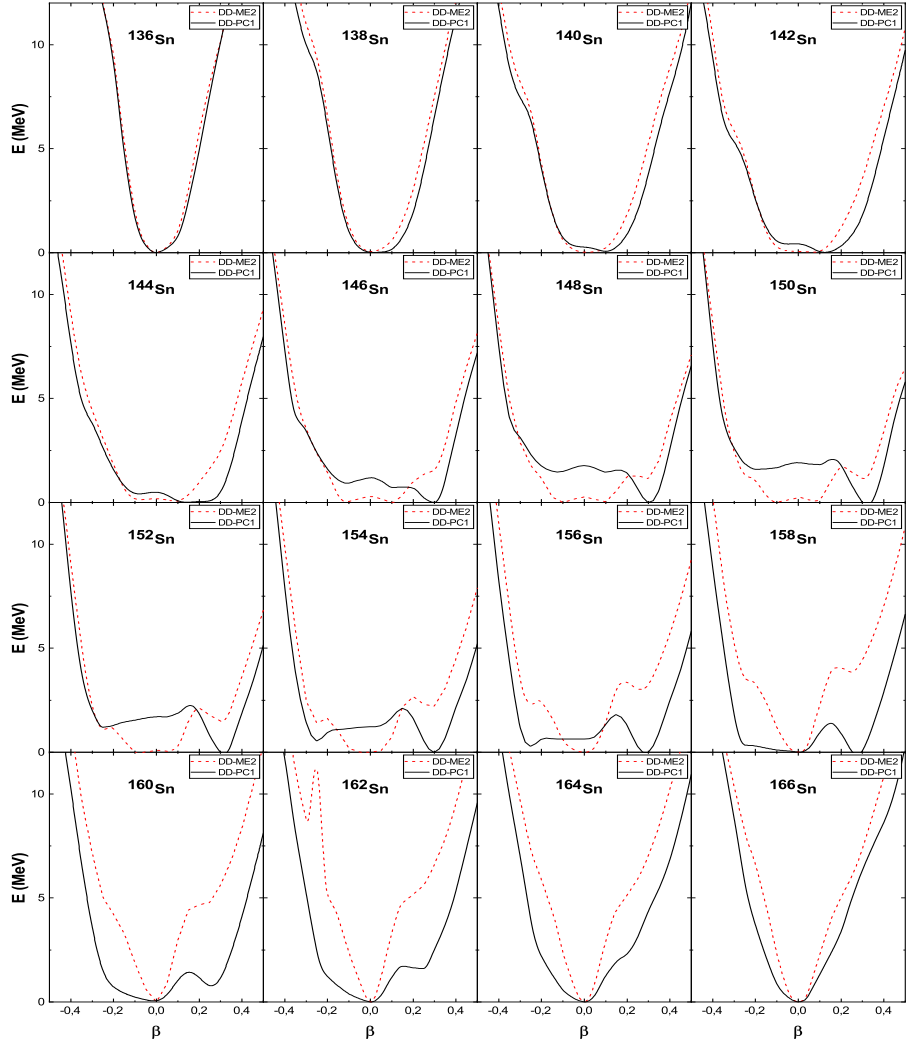}
	\caption{(Color online) The total energy curves of 16 Sn isotopes obtained within CDFT framework with DD-ME2 and DD-PC1 sets as functions of the mass quadrupole deformation parameters, $\beta_2$.}
	\label{16beta}
\end{figure}

We display in Fig.~\ref{16beta} the total energy curves of 16 Sn isotopes, $^{136-166}$Sn, as functions of   $\beta_2$ and obtained by using the density-dependent effective interactions DD-ME2 and DD-PC1. This figure provides us a clear idea about how deformation evolves in this isotopic chain. Prolate shapes correspond to $\beta >0$  while oblate
shapes correspond to $\beta <0$.

%%%%%%%%%%%%%%%%%%%%%%%%%%%%%%%%%%%%%%%%%%%%%%%%%%%%%
%%%%%%%%%%%%%%%%%%%%%%%%%%%%%%%%%%%%%%%%%%%%%%%%%%%%%

From Fig.~\ref{16beta}, it is seen that nuclei with $A < 140$ are
spherical. At A=142, very shallow deformation degenerate minima (oblate
 and prolate) appear. The next isotopes develop
 a more well-deformed prolate minimum, which reaches $\beta \approx 0.3$ for $^{150}$Sn in DD-PC1 calculations. 
 As the mass number increases, the two minima gradually disappear until A = 166 where we get a sharp single minimum, which confirms the spherical shape at this isotope. The total energy curves of $^{136-166}$Sn isotopes shown in this figure are in good agreement with the results shown in Fig.~\ref{beta}.

\subsection{Neutron, Proton and Charge radii}

\begin{figure}[ht]								   
	\centering \includegraphics[scale=0.6]{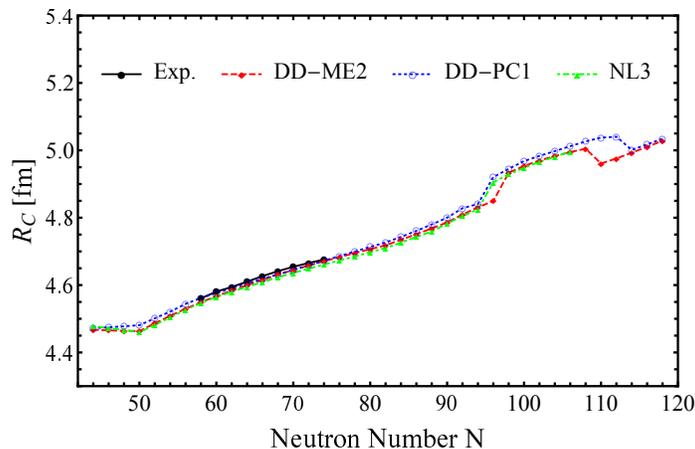}
	\caption{(Color online) The charge radii obtained by our DD-ME2 and DD-PC1 calculations and compared with the available experimental data~\cite{Vries1987,Fricke1995,angeli2004} and predictions of RMF(NL3)~\cite{lalazissis1999}}
	\label{Rc}
\end{figure}

\begin{figure}[!htb]
	\minipage{0.48\textwidth}
	\centering \includegraphics[scale=0.4]{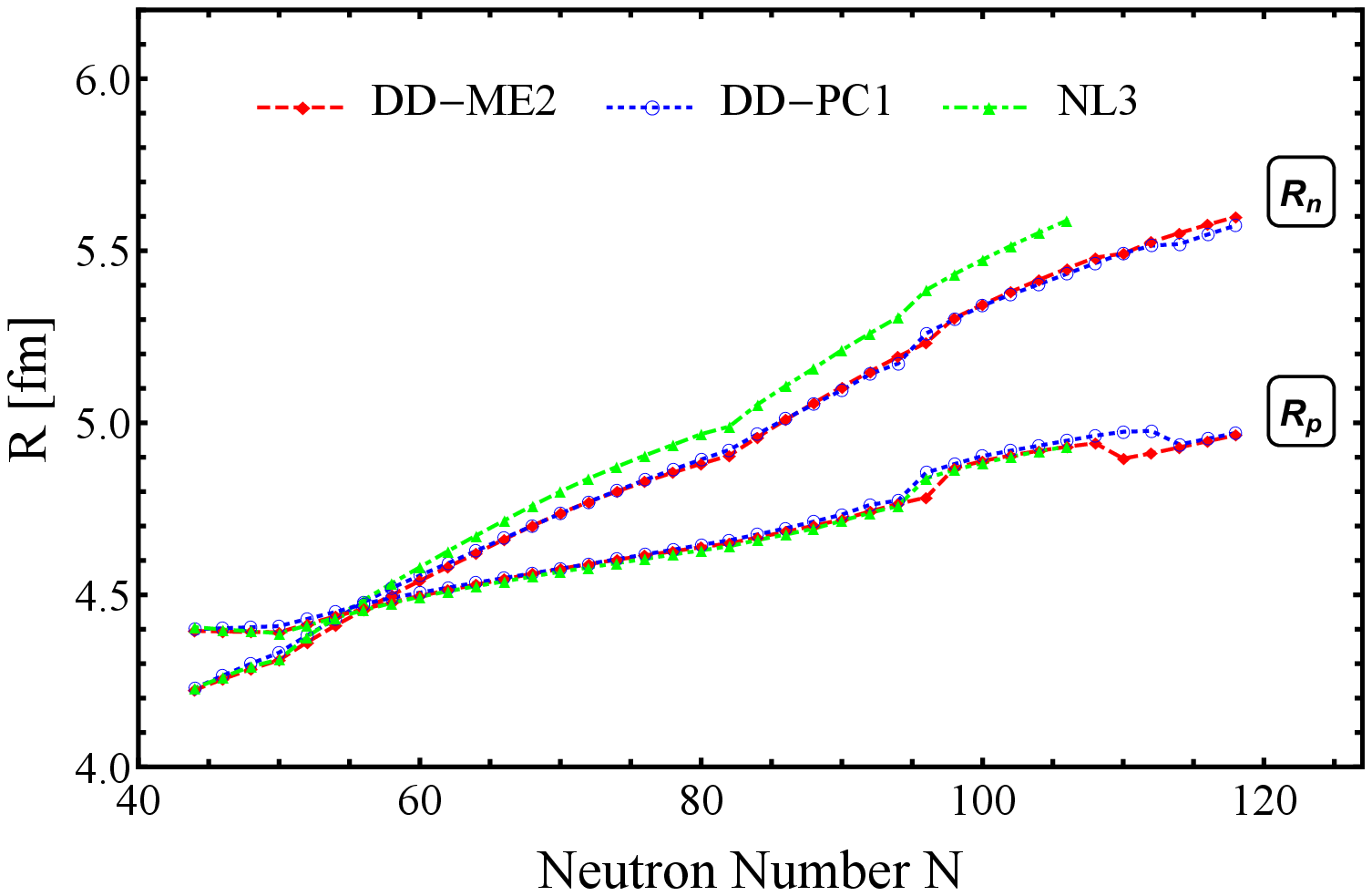}
	\endminipage\hfill
	\minipage{0.48\textwidth}
	\centering \includegraphics[scale=0.4]{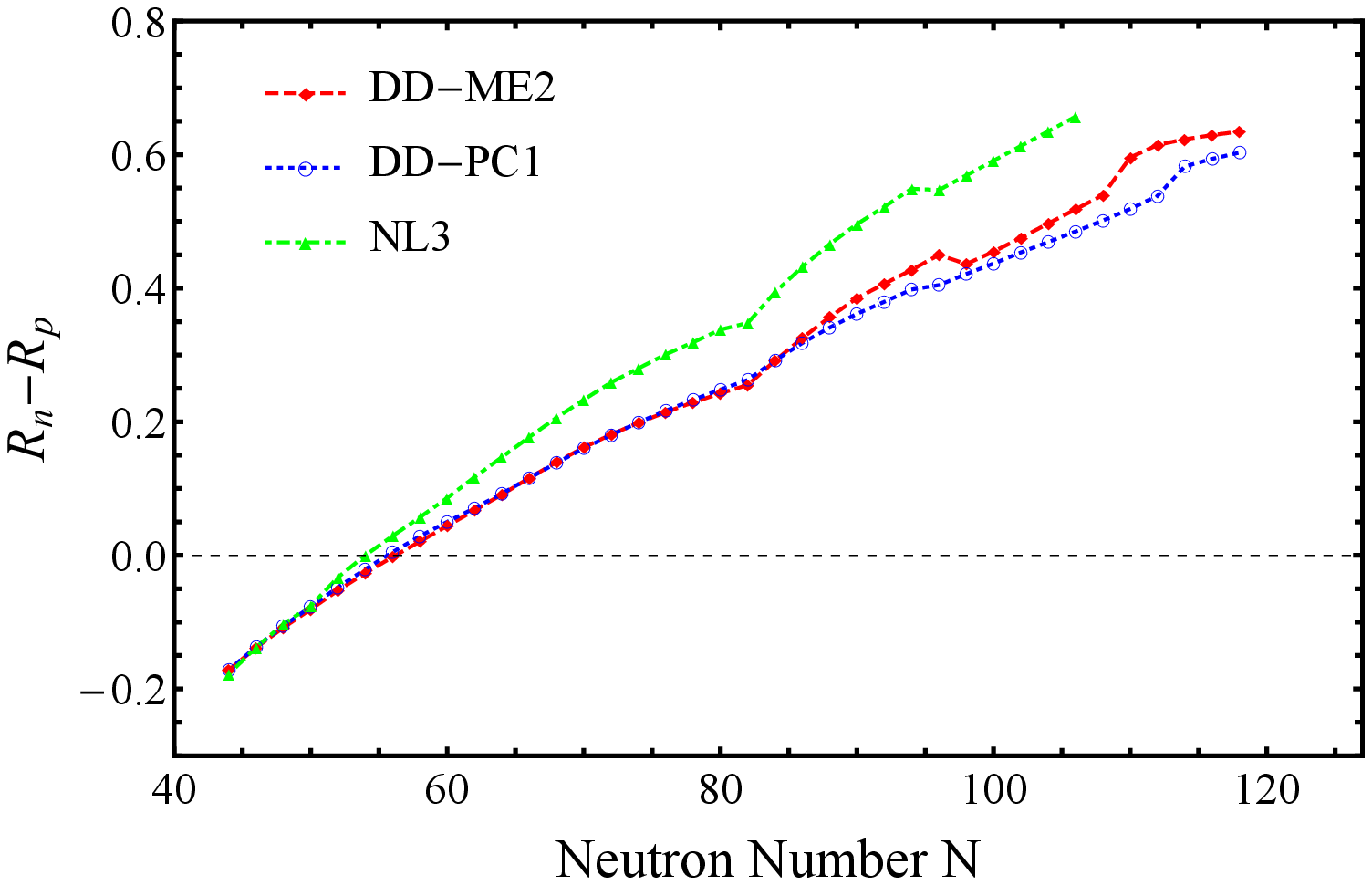}
	\endminipage\hfill
	\caption{(Color online) The neutron and proton radii of Sn isotopes (left panel) and the neutron skin thicknesses ($\Delta R$ = $R_n$ - $R_p$)(right panel).}
	\label{RnRp}
\end{figure}

In this subsection, we present the nuclear radii evaluated using different interactions. Fig.~\ref{Rc} shows root mean square charge radii and Fig.~\ref{RnRp} shows neutron and proton radii for all tin isotopes.

The charge radius, $R_c$, is related to the proton radius, $R_p$, by:
\begin{equation}
	R_c^2=R_p^2+0.64~(fm^2)
	\label{eqRc}
\end{equation} 
where the factor 0.64 in Eq.~(\ref{eqRc}) is a correction due to the finite size of the proton.

Fig.~\ref{Rc} shows the charge radii calculated within CDFT with DD-ME2 and DD-PC1 in comparison with the available experimental data~\cite{Vries1987,Fricke1995,angeli2004} and with predictions of RMF(NL3)~\cite{lalazissis1999}. Excellent agreement between theory and experiment can be clearly seen in Fig.~\ref{Rc}. 
The small jump in $R_c$ starting from $^{146}$Sn up to $^{162}$Sn seems to be due to the deformation effect as it was previously explained in Ref~\cite{geng2004}. Indeed, it becomes much clearer if we take a look at Fig.~\ref{beta}: $\beta_2$ changes, in the case of DD-PC1 for example, from 0.0 for $^{144}$Sn to 0.284 for $^{146}$Sn, and it changes from 0.223 for $^{162}$Sn to 0.0 for $^{164}$Sn.

%%%%%%%%%%%%%%%%%%%%%%%%%%%%%%%%%%%%%%%%%%%%%%%%%%%%%
%%%%%%%%%%%%%%%%%%%%%%%%%%%%%%%%%%%%%%%%%%%%%%%%%%%%%

%%%%%%%%%%%%%%%%%%%%%%%%%%%%%%%%%%%%%%%%%%%%%%%%%%%%%
%%%%%%%%%%%%%%%%%%%%%%%%%%%%%%%%%%%%%%%%%%%%%%%%%%%%%

Fig. \ref{RnRp} shows the calculated neutron and proton radii of tin isotopes obtained by using DD-ME2 and DD-PC1. The results of RMF model with NL3 functional are also shown for comparison. R$_n$ and R$_p$ are plotted together in the same figure in order to see the differences between them easily.

In Fig.~\ref{RnRp} (left panel), we see that the neutron rms radii obtained by using DD-ME2 and DD-PC1 are almost equal
throughout the isotopic chain, but the results of  NL3 are overestimated.
The proton rms radii by both formalisms, CDFT(DD-ME2 and DD-PC1) and RMF(NL3), are almost the same.

We also notice from the right panel of Fig.~\ref{RnRp} that the difference between the rms radii of neutrons and protons ($\Delta R = R_n - R_p$) increases monotonously by increasing the neutron number in favor of developing a neutron skin, which reaches 0.6 fm for $^{168}$Sn.

\section{Conclusion}
\label{Conclusion}

In this work, the covariant density functional theory has been employed to investigate the ground state properties of even-even tin isotopes, $^{94-168}$Sn, from the proton drip line to the neutron drip line, by using 
two of the best available functionals: The density-dependent meson-exchange DD-ME2  and the density-dependent point-coupling DD-PC1.
Binding energy,  two-neutron separation energy (S$_{2n}$),  two-neutron shell gap ($\delta_{2n}$), rms-radii for protons and neutrons, pairing energy and quadrupole deformation have been calculated. Some indications on shape phase transition for Sn isotopic chain were given. Our results are in good agreement with the experimental data. 
DD-ME2 functional seems to be more accurate than DD-PC1 and RMF(NL3).
A strong shell closure is observed at N=50 and N = 82. The neutron skin in our calculation reaches 0.6 fm for $^{168}$Sn.

\section*{Acknowledgment}
We are very grateful to T. Nik\v{s}i\'{c}, N. Paar, D. Vretenar, and
P. Ring for making available the DIRHB package of FORTRAN
computer codes.


\section*{References}

\begin{thebibliography}{99}
	\bibitem{moller1995}
	P. Moller, et al. "Nuclear ground-state masses and deformations." Atomic data and nuclear data tables 59 185-381 (1995).
	
	\bibitem{samanta2002}
	C. Samanta, and S. Adhikari, Physical Review C 65.3,  037301 (2002).
	
	\bibitem{skyrme1956}
	T. H. R. Skyrme,  "CVII. The nuclear surface." Philosophical Magazine 1.11, 1043-1054 (1956).
	
	\bibitem{decharg1980}
	J. Dechargé and D. Gogny, Phys. Rev. C 21, 1568 (1980)." Phys. Rev. C 21 1568 (1980).
	
	\bibitem{bassem2015}
	Y. El Bassem and M. Oulne, Int. J. Mod. Phys. E 24, 1550073 (2015).
	
	\bibitem{bassem2016}
	Y. El Bassem and M. Oulne, Nucl. Phys. A 957 22 (2017).
	
	\bibitem{bassem2017}
	Y. El Bassem and M. Oulne, Int. J. Mod. Phys. E 26, 1750084 (2017).
	
	\bibitem{walecka1974}
	J. D. Walecka, Ann. Phys. (N. Y.) 83, 491 (1974).
	
	\bibitem{reinhard1989}
	P. G. Reinhard, Rep. Prog. Phys. 52, 439 (1989).
	
	\bibitem{meng1998}
	J. Meng and P. Ring, Phys. Rev. Lett. 80, 460 (1998).
	
	\bibitem{zhou2003}
	S. G. Zhou, J. Meng, and P. Ring, Phys. Rev. Lett. 91, 262501 (2003).
	
	\bibitem{meng1999}
	J. Meng, K. Sugawara-Tanabe, S. Yamaji, and A. Arima, Phys.
	Rev. C 59, 154 (1999).
	
	\bibitem{geng2004}
	L. S. Geng, H. Toki, and J. Meng. Modern Physics Letters A 19, 2171  (2004).
	
	\bibitem{ring1996}
	P. Ring, Prog. Part. Nucl. Phys. 37, 193 (1996).
	
	\bibitem{meng1996}
	J. Meng and P. Ring, Phys. Rev. Lett. 77, 3963 (1996).
	
	\bibitem{ginocchio1997}
	J. N. Ginocchio, Phys. Rev. Lett. 78, 436 (1997).
	
	\bibitem{kohn1965}
	W. Kohn and L. J. Sham, Phys. Rev. 137, A1697 (1965).
	
	\bibitem{kohn1965(2)}
	W. Kohn and L. J. Sham, Phys. Rev. 140, A1133 (1965).
	
	\bibitem{niki2008}
	T. Nik\v{s}i\'{c}, D. Vretenar and P. Ring, Phys. Rev. C 78, 034318 (2008).
	
	\bibitem{niki2014}
	T. Nik\v{s}i\'{c}, N. Paar, D. Vretenar, and P. Ring, Comp. Phys. Commun. 185, 1808 (2014).
	
	\bibitem{lalazissis2005}
	G. A. Lalazissis, T. Nik\v{s}i\'{c}, D. Vretenar and P. Ring, Phys. Rev. C 71, 024312 (2005).
	
	\bibitem{rocamaza2011}
	X. Roca-Maza, X. Vi\~{n}as, M. Centelles, P. Ring and P. Schuck, Phys. Rev. C 84,
	054309 (2011).
	
	\bibitem{niki2002}
	T. Nik\v{s}i\'{c}, D.Vretenar, P. Finelli and P. Ring, Phys. Rev. C 66 (2002) 024306.
	
	\bibitem{meng2015}
	J. Meng and S. G. Zhou, J. Phys. G 42, 093101 (2015).
	
	\bibitem{matev2007}
	M. Matev, A. V. Afanasjev, J. Dobaczewski, G. A. Lalazissis and W. Nazarewicz,
	Phys. Rev. C 76, 034304 (2007).
	
	\bibitem{afanasjev2008}
	A. V. Afanasjev, Phys. Rev. C 78, 054303 (2008).
	
	\bibitem{abusara2012}
	H. Abusara, A. V. Afanasjev and P. Ring, Phys. Rev. C 85, 024314 (2012).
	
	\bibitem{agbemava2015}
	S. E. Agbemava, A. V. Afanasjev, T. Nakatsukasa and P. Ring, Phys. Rev. C
	92, 054310 (2015).
	
	\bibitem{agbemava2017}
	S. E. Agbemava, A. V. Afanasjev, D. Ray and P. Ring, Phys. Rev. C 95, 054324 (2017).
	
	\bibitem{Zhao2014}
	P. W. Zhao, S. Q. Zhang, and J. Meng, Phys. Rev. C 89 011301 (2014).	

	\bibitem{Zhang2014}
	P. Zhang, B. Qi, and S. Y. Wang,  Phys. Rev. C 89, 047302 (2014).
	
	\bibitem{karim2015}
	A. Karim and S. Ahmad, Phys. Rev. C, 92, 064608 (2015).
	
	\bibitem{afanasjev2010}
	A. V. Afanasjev and H. Abusara, Phys. Rev. C 81, 014309 (2010).
	
	\bibitem{typel1999}
	S. Typel and H. H. Wolter, Nucl. Phys. A 656, 331 (1999).
	
	
	\bibitem{lalazissis2009}
	G. A. Lalazissis, S. Karatzikos, R. Fossion, D. Pe˜na Arteaga,
	A. V. Afanasjev, and P. Ring, Phys. Lett. B 671, 36 (2009).
	
	
	\bibitem{gambhir1990}
	Y. K. Gambhir, P. Ring, and A. Thimet, Ann. Phys. (NY) 198, 132 (1990).
	
	
	\bibitem{boguta1977}
	J. Boguta and R. Bodmer, Nucl. Phys. A 292, 413 (1977).
	
	\bibitem{Pannert1987}
	W. Pannert, P. Ring, and J. Boguta, Physical review letters, 59, 2420 (1987).
	
	
	\bibitem{reinhard1986}
	P. G. Reihard, M. Rufa, J.Maruhn,W. Greiner, and J. Friedrich,
	Z. Phys. A-Atomic Nuclei 323, 13 (1986).
	
	\bibitem{lalazissis1997}
	G. A. Lalazissis, J.König, P. Ring, Phys. Rev. C 55, 540 (1997).
	
	\bibitem{toddrutel2005}
	B. G. Todd-Rutel and J. Piekarewicz, Phys. Rev. Lett. 95, 122501 (2005).
	
	\bibitem{DD-ME2}
	G. A. Lalazissis, T. Nik\v{s}i\'{c}, D. Vretenar, and P. Ring, Phys.
	Rev. C 71, 024312 (2005).
		
	\bibitem{nikolaus1992}
	B. A. Nikolaus, T. Hoch, and D. G. Madland, Phys. Rev. C 46, 1757 (1992).
	
	\bibitem{lalazissis1999}
	G.A. Lalazissis, S. Raman, P. Ring, At. Data Nucl. Data Tables 71, 1–40 (1999).
	
	\bibitem{Audi2003}
	G. Audi, A. H. Wapstra, and C. Thibault. Nuclear physics A 729, 337 (2003).
	
	\bibitem{wang2012}
	M. Wang, G. Audi, A.H. Wapstra, et al., The Ame2012 atomic mass evaluation, Chin. Phys. C 36, 1603 (2012).
	
	\bibitem{beta_experimenta_data}
	S. Raman, C. W. Nestor Jr, and P. Tikkanen, Atomic Data and Nuclear Data Tables, 78, 1-128 (2001).
	
	
	\bibitem{Zongqiang2012}
	S. Zongqiang, and R. Zhongzhou. Plasma Science and Technology 14, 534 (2012).

	
	\bibitem{Vries1987}
	H. De Vries, C. W. De Jager, and C. De Vries, At. Data Nucl. Data Tables 36, 495 (1987)
	
	\bibitem{Fricke1995}	
	 G. Fricke, et al., At. Data Nucl. Data Tables 60, 177 (1995).
	
	\bibitem{angeli2004}
	I. Angeli, At. Data Nucl. Data Tables 87, 185–206 (2004).
	
	
	
	
\end{thebibliography}
\end{document}